\documentclass[10pt,apjl]{emulateapj}
\usepackage[dvips]{color}

\usepackage{apjfonts}

\newcommand{\teff}{$T_{\rm eff}$} 
\newcommand{\logg}{$\log g$} 
\newcommand{\kms}{km s$^{-1}$}
\newcommand{\fei}{Fe\,{\sc i}}

\shorttitle{The Most Metal-Poor Stars III.}
\shortauthors{Yong et al.}

\begin{document}

\title{THE MOST METAL-POOR STARS. III. THE METALLICITY DISTRIBUTION FUNCTION 
AND CEMP FRACTION\altaffilmark{1,2,3}}

\author{
DAVID YONG\altaffilmark{4},
JOHN E.\ NORRIS\altaffilmark{4}, 
M.\ S.\ BESSELL\altaffilmark{4},
N.\ CHRISTLIEB\altaffilmark{5},
M.\ ASPLUND\altaffilmark{4,6},
TIMOTHY C.\ BEERS\altaffilmark{7,8},
P.\ S.\ BARKLEM\altaffilmark{9}, 
ANNA FREBEL\altaffilmark{10}, AND 
S.\ G.\ RYAN\altaffilmark{11}
}

\altaffiltext{1}{This paper includes data gathered with the 6.5 meter 
Magellan Telescopes located at Las Campanas Observatory, Chile.} 

\altaffiltext{2}{Some of the data presented herein were obtained at the 
W.\ M.\ Keck Observatory, which is operated as a scientific partnership 
among the California Institute of Technology, the University of California 
and the National Aeronautics and Space Administration. The Observatory was 
made possible by the generous financial support of the W.\ M.\ Keck 
Foundation.}

\altaffiltext{3}{Based on observations collected at the 
European Organisation for Astronomical Research in the 
Southern Hemisphere, Chile (proposal 281.D-5015).}

\altaffiltext{4}{Research School of Astronomy and Astrophysics, The
Australian National University, Weston, ACT 2611, Australia;
yong@mso.anu.edu.au, jen@mso.anu.edu.au, bessell@mso.anu.edu.au,
martin@mso.anu.edu.au}

\altaffiltext{5}{Zentrum f\"ur Astronomie der Universit\"at
Heidelberg, Landessternwarte, K{\"o}nigstuhl 12, D-69117 Heidelberg,
Germany; n.christlieb@lsw.uni-heidelberg.de}

\altaffiltext{6}{Max-Planck Institute for Astrophysics, 
Karl-Schwarzschild Str. 1, 85741, Garching, Germany} 

\altaffiltext{7}{National Optical Astronomy Observatory, Tucson, AZ 85719} 

\altaffiltext{8}{Department of Physics \& Astronomy and JINA: Joint
Institute for Nuclear Astrophysics, Michigan State University,
E. Lansing, MI 48824, USA; beers@pa.msu.edu}

\altaffiltext{9}{Department of Physics and Astronomy, Uppsala
University, Box 515, 75120 Uppsala, Sweden;
paul.barklem@physics.uu.se}

\altaffiltext{10}{Massachusetts Institute of Technology, 
Kavli Institute for Astrophysics and Space Research, Cambridge, MA 02139, USA;
afrebel@mit.edu} 

\altaffiltext{11}{Centre for Astrophysics Research, School of Physics,
Astronomy \& Mathematics, University of Hertfordshire, College Lane,
Hatfield, Hertfordshire, AL10 9AB, UK; s.g.ryan@herts.ac.uk}

\newpage
\begin{abstract}

We examine the metallicity distribution function (MDF) and fraction of
carbon-enhanced metal-poor (CEMP) stars in a sample that includes 86
stars with [Fe/H]~$\le$~$-$3.0, based on high-resolution, high-S/N
spectroscopy, of which some 32 objects lie below [Fe/H]~=~$-$3.5.  
After accounting for the completeness function, 
the ``corrected'' MDF does not exhibit the sudden drop at [Fe/H] = $-$3.6 that was found in
recent samples of dwarfs and giants from the Hamburg/ESO survey.
Rather, the MDF decreases smoothly down to 
[Fe/H] = $-$4.1. 
Similar results are obtained from the ``raw'' MDF. 
We find the fraction of CEMP objects below
[Fe/H]~=~$-$3.0 is 23 $\pm$ 6\% and 32 $\pm$ 8\% when adopting
the \citeauthor{beers05} and \citeauthor{aoki07} CEMP definitions,
respectively. The former value is in fair agreement with some previous
measurements, which adopt the \citeauthor{beers05} criterion.

\end{abstract}

\keywords{Cosmology: Early Universe, Galaxy: Formation, Galaxy: Halo, 
Nucleosynthesis, Abundances, Stars: Abundances}

\section{INTRODUCTION}
\label{sec:intro}

Metal-poor stars provide critical information on the earliest phases 
of Galactic formation (see e.g., the reviews by \citealt{beers05} and 
\citealt{frebel11}). 
Their chemical abundances shed light upon 
the nature of the first stars to have formed in the Universe,  
and the nucleosynthesis which seeded all subsequent generations of 
stars.  

This is the third paper in our series, which focuses upon the discovery of,
and high-resolution, high signal-to-noise ratio (S/N) spectroscopic analysis of,
the most metal-poor stars. 
Here we explore two key issues:
the metallicity distribution function (MDF) and the fraction of
carbon-enhanced metal-poor (CEMP)\footnote{Initially
defined as stars with [C/Fe] $\ge$ +1.0 and [Fe/H] $\le$ $-$2.0
\citep{beers05}.} stars at lowest metallicities.

Any model purporting to explain the formation and evolution of our
Galaxy must be able to reproduce the observed MDF.  The ingredients of
such models include the initial mass function (IMF), nucleosynthetic
yields, and inflow or outflow of gas.  Observations of the MDF can
constrain these initial conditions and physical processes.  Since the
early work by \citet{hartwick76}, measurements of the MDF involve
increasing numbers of stars with more accurate metallicity
measurements (see e.g., \citealt{laird88,ryan91}). One of the basic
predictions of Hartwick's Simple Model of Galactic Chemical
Enrichment is that the number of stars having abundance less than
a given metallicity should decrease by a factor of ten for each
factor of ten decrease in metallicity{\footnote{While a number of chemical
evolution models (e.g., \citealt{kobayashi06},
\citealt{karlsson06}, \citealt{salvadori07}, \citealt{prantzos08},
and \citealt{cescutti10}) have improved upon the one-zone
closed-box Hartwick model, the general behavior remains largely
unchanged.}. \citet{norris99} presented
observational support for this suggestion, down to [Fe/H]
$\sim-4.0$, below which it appeared to be no longer valid. More
recently, \citet{schorck09} and \citet{li10} presented MDFs of the
Galactic halo using 1638 giant and 617 dwarf stars, respectively,
from the Hamburg/ESO Survey (HES: \citealt{hes}).  Below [Fe/H] =
$-$2.5, the MDFs for dwarfs and giants were in excellent agreement.
A prominent feature of both MDFs was the apparent lack of stars more
metal-poor than [Fe/H] = $-$3.6. While a handful of such stars are
known, the sharp cutoff in the MDF has important implications for
the critical metallicity above which low-mass star formation is
possible (e.g., \citealt{salvadori07}).  More detailed studies of
the MDF, and in particular the low-metallicity tail, are necessary
to confirm and constrain the star formation modes of the first stars
(e.g., \citealt{bromm04}).

The HK survey \citep{beers85,beers92} revealed that there is a large
fraction of metal-poor stars with unusually strong CH $G$-bands
indicating high C abundances. With the addition of numerous metal-poor
stars found in the HES, the CEMP fraction at low metallicity has been
confirmed and quantified, with estimates ranging from 9\%
\citep{frebel06} to > 21\% \citep{lucatello06}.  These numbers are
considerably larger than the fraction of C-rich objects at higher
metallicity, the so-called CH and Ba stars, which account for only
$\sim$ 1\% of the population.  The fraction is even
larger at lowest metallicity: below [Fe/H] $<$ --4.5, 75\% of 
the four known stars
belong to the CEMP class \citep{norris07,caffau11}.  To explain these
large fractions, several studies argue that adjustments to the IMF are
necessary (e.g., \citealt{lucatello05,komiya07,izzard09}).  
\citet{carollo12} offer an alternative interpretation for
the increase of the CEMP fraction they observe in the range $-$3.0
$<$ [Fe/H] $<$ $-$1.5 in terms of a dependence of CEMP fraction on
height above the Galactic plane. In their most metal-poor bin at
[Fe/H] $\sim$ --2.7, they report C-rich fractions of 20\% and 30\%
for their inner- and outer-halo components, respectively (see their
Figure 15). In their view, this can be accounted for by the presence
of different carbon-production mechanisms (some not involving the
presence of AGB nucleosynthesis) that have operated in the inner-
and outer-halo populations. 

An understanding of the CEMP stars is complicated by the fact that
they do not form a homogeneous group: \citet{beers05} define several
CEMP subclasses (all of which have [C/Fe] $>$ +1.0) as follows: (i) 
CEMP-r -- [Eu/Fe] $>$ +1.0; (ii) CEMP-s -- [Ba/Fe] $>$ +1.0 and
[Ba/Eu] $>$ + 0.5; (iii) CEMP r/s -- 0.0 $<$ [Ba/Eu] $<$ +0.5; and
CEMP-no -- [Ba/Fe] $<$ 0.0.  \citet{aoki10} shows that below
[Fe/H] = --3.0, the CEMP stars are principally (90\%) CEMP-no stars,
while for [Fe/H] $>$ --3.0, the CEMP-s class predominates.
These differences lie outside the scope of
the present paper.  Here we seek to constrain only the fraction of
CEMP stars at lowest abundance, [Fe/H] $<$ --3.0, and to compare the
results with the fractions determined at higher abundances.  In Paper IV
(Norris et al.\ 2012b) we shall address the nature of the CEMP-no
stars, which comprise the large majority of CEMP stars in our
extremely metal-poor sample.

\section{OBSERVATIONS AND ANALYSIS}
\label{sec:obs}

In Norris et al.\ (2012a; Paper I), we presented high-resolution spectroscopic
observations of 38 extremely metal-poor stars ([Fe/H] $<$ $-$3.0; 34 newly
discovered), obtained using the Keck, Magellan, and VLT telescopes, including
the discovery and sample selection, equivalent-width measurements, radial
velocities, and line list. 
In Paper I, we also described the
temperature scale, which consists of spectrophotometry and Balmer-line
analysis.  In addition to the 38 program stars, we selected 207 stars
from the SAGA database \citep{saga} (queried on 2 Feb 2010), and performed a homogeneous
re-analysis of this literature sample.  
All stars were analyzed using
the NEWODF grid of ATLAS9 model atmospheres \citep{castelli03}, and the
2011 version of the stellar line-analysis program MOOG \citep{moog}, 
which includes proper treatment of continuum scattering
\citep{sobeck11}.  They thus have effective temperatures, surface
gravities, microturbulent velocities, log $gf$ values, solar
abundances \citep{asplund09}, and therefore metallicities, [Fe/H], all on
the same scale.  

The literature sample was reduced from 207 to 152 stars by 
(i) discarding stars with fewer than 14 \fei\ lines (the minimum 
number of \fei\ lines measured in our program stars), 
(ii) removing literature stars included in the program-star sample, and 
(iii) averaging the results of stars having multiple analyses into a single 
set of abundances. 
Thus, the final combined sample consists of 190 stars (38 program 
stars and 152 literature stars). 
Full details regarding the analysis are
presented in Yong et al.\ (2012, Paper II). 

\section{RESULTS}
\label{sec:abund}

\subsection{Selection Biases}\label{sec:biases}

In Table \ref{tab:param}, we present data, 
based on our high-resolution analyses, for the 
86\footnote{For nine program stars, we could not determine
whether they were dwarfs or subgiants. For the subset of those stars
included in this paper, we present the results for both cases in
Table \ref{tab:param}. In all figures, unless noted otherwise, we
adopt the average [Fe/H] and [X/Fe] from the dwarf and subgiant 
analyses for these stars. For the nine objects, the
average differences are $\langle$[Fe/H]$_{\rm dwarf}$
$-$ [Fe/H]$_{\rm subgiant}\rangle$ = 0.02 $\pm$ 0.01 dex
($\sigma$ = 0.03) and $\langle$[X/Fe]$_{\rm dwarf}$ $-$
[X/Fe]$_{\rm subgiant}\rangle$ = 0.05 $\pm$ 0.02 dex ($\sigma$ =
0.17),  where X refers to the 14 species (from Na to Ba) 
measured in Paper II. For C, while the differences are larger, 
[C/Fe]$_{\rm subgiant}$ $-$ [C/Fe]$_{\rm dwarf}$ = 0.23 $\pm$ 0.05 dex  
($\sigma$ = 0.13), 
the CEMP classifications do not depend on whether we adopt the dwarf or 
subgiant value.} stars in our
collective sample that have [Fe/H] $\le$ $-$3.0; of these,
32 have [Fe/H] $\le$ $-$3.5, while there are nine with [Fe/H] $\le$
$-$4.0. We stress again that these metallicities are on our
homogeneous system of \teff, $\log g$, $\xi_t$, $\log gf$ values, and
solar abundances.  
These are the most metal-poor stars known in our Galaxy, 
and allow us to address below the key issues of the MDF
and CEMP fraction.

\begin{deluxetable*}{lrrrrrrrrrr}
\tabletypesize{\scriptsize}
\tablecolumns{9} 
\tablewidth{0pc} 
\tablecaption{Stellar Parameters and Carbon Abundance \label{tab:param}}
\tablehead{ 
\colhead{Star} & 
\colhead{RA2000\tablenotemark{a}} &
\colhead{DEC2000\tablenotemark{a}} &
\colhead{\teff} & 
\colhead{$\log g$} & 
\colhead{$\xi_t$} & 
\colhead{[M/H]$_{\rm model}$} & 
\colhead{[Fe/H]$_{\rm derived}$} & 
\colhead{[C/Fe]\tablenotemark{b}} & 
\colhead{C-rich\tablenotemark{c}} & 
\colhead{Source} \\
\colhead{} & 
\colhead{} &
\colhead{} &
\colhead{(K)} & 
\colhead{(cgs)} & 
\colhead{(\kms)} & 
\colhead{} & 
\colhead{} & 
\colhead{} \\
\colhead{(1)} & 
\colhead{(2)} &
\colhead{(3)} &
\colhead{(4)} & 
\colhead{(5)} & 
\colhead{(6)} & 
\colhead{(7)} & 
\colhead{(8)} & 
\colhead{(9)} & 
\colhead{(10)} &
\colhead{(11)} 
}
\startdata 
CS29527-015 &  00 29 10.7 & $-$19 10 07.2 & 6577 & 3.89 & 1.9 & $-$3.3 & $-$3.32 & 1.18 & 1 &   5 \\
CS30339-069 &  00 30 16.0 & $-$35 56 51.2 & 6326 & 3.79 & 1.4 & $-$3.1 & $-$3.05 & 0.56 & 0 &   5 \\
CS29497-034 &  00 41 39.8 & $-$26 18 54.4 & 4983 & 1.96 & 2.0 & $-$3.0 & $-$3.00 & 2.72 & 1 &   4 \\
HD4306 &  00 45 27.2 & $-$09 32 39.9 & 4854 & 1.61 & 1.6 & $-$3.1 & $-$3.04 & 0.11 & 0 &  12 \\
CD-38 245 &  00 46 36.2 & $-$37 39 33.5 & 4857 & 1.54 & 2.2 & $-$4.2 & $-$4.15 & $<-$0.33 & 0 &   7 \\
HE0049-3948 &  00 52 13.4 & $-$39 32 36.9 & 6466 & 3.78 & 0.8 & $-$3.7 & $-$3.68 & <1.81 & 0 &   1 \\
HE0057-5959 &  00 59 54.0 & $-$59 43 29.9 & 5257 & 2.65 & 1.5 & $-$4.1 & $-$4.08 & 0.86 & 1 &   1 \\
HE0102-1213 &  01 05 28.0 & $-$11 57 31.1 & 6100 & 3.65 & 1.5 & $-$3.3 & $-$3.28 & <1.31 & 0 &   1 \\
CS22183-031 &  01 09 05.1 & $-$04 43 21.1 & 5202 & 2.54 & 1.1 & $-$3.2 & $-$3.17 & 0.42 & 0 &  12 \\
HE0107-5240 &  01 09 29.2 & $-$52 24 34.2 & 5100 & 2.20 & 2.2 & $-$5.3 & $-$5.54 & 3.85 & 1 &   8 
\enddata

\tablenotetext{a}{Coordinates are on the 2MASS system \citep{2mass}}
\tablenotetext{b}{For literature stars, [C/Fe] is the (average) value from the reference(s).}
\tablenotetext{c}{We adopt the \citet{aoki07} CEMP definition.} 
\tablenotetext{d}{This analysis assumes the star is a dwarf.} 
\tablenotetext{e}{This analysis assumes the star is a subgiant.} 

\tablerefs{
1 = This study; 
2 = \citet{aoki02}; 
3 = \citet{aoki06}; 
4 = \citet{aoki07}; 
5 = \citet{bonifacio07,bonifacio09}; 
6 = \citet{carretta02,cohen02}; 
7 = \citet{cayrel04,spite05}; 
8 = \citet{christlieb04}; 
9 = \citet{cohen06}; 
10 = \citet{cohen08}; 
11 = \citet{frebel07}; 
12 = \citet{honda04}; 
13 = \citet{lai08}; 
14 = \citet{norris01}; 
15 = \citet{norris07}
\\
\\
Note. Table 1 is published in its entirety in the electronic edition of The Astrophysical Journal. A portion is shown here for guidance regarding its form and content.
}

\end{deluxetable*}

Before continuing, we comment on the completeness function and
selection biases of the sample.  The HES is complete for
metallicities below [Fe/H] = $-$3.0 \citep{schorck09,li10}.  
To estimate the completeness, \citet{schorck09} and \citet{li10} 
used the Simple Model to generate a metallicity distribution function 
and then applied their selection criteria to obtain the MDF which would 
have been observed in the HES (see Section 6 in \citealt{schorck09}, and 
Section 3.4 in \citealt{li10} for further details). 
From
Paper I, we can compute the completeness function for the $\sim30$ HES
candidates having high-resolution, high-S/N spectra discovered in
that work.  First, we use a linear transformation to place the 
medium-resolution metallicities, [Fe/H]$_{\rm K}$, from Paper I onto the
high-resolution abundance scale, [Fe/H].  We then compare the number
of HES stars observed at high resolution with the total number of HES
stars observed at medium resolution, and from which the stars observed
at high resolution were selected, as a function of [Fe/H]. We use this
ratio to correct the MDFs in the following subsection.  In a similar
manner, we are able to determine the completeness function for the
$\sim$ 50 HK-survey stars in our extended sample, by using material in the
medium-resolution HK database maintained by T.\ C.\ B.

\subsection{The Metallicity Distribution Function (MDF)}

Our MDFs are presented in Figure \ref{fig:mdf}\footnote{ All figures were
generated using the full sample, presented in Paper II.}, where in
the left panels the scale of the ordinate is linear and for those on the
right it is logarithmic.  The two uppermost panels each contain MDFs
constructed from the raw data for the 38 program stars and the total
sample of 190 objects. 
We use generalized histograms, in which each data point is replaced by
a Gaussian of width $\sigma$ = 0.30\footnote{We regard our typical
uncertainty in [Fe/H] to be 0.15 dex, rather than 0.30 dex.  Given
our still relatively limited sample size, using $\sigma$ = 0.15 dex
produces spurious structure in our MDF. None of our conclusions
depend upon our choice of $\sigma$ in constructing the MDF.} dex.
The Gaussians are then summed to produce a realistically smoothed
histogram.

\begin{figure*}[t!]
\epsscale{0.8}
\plotone{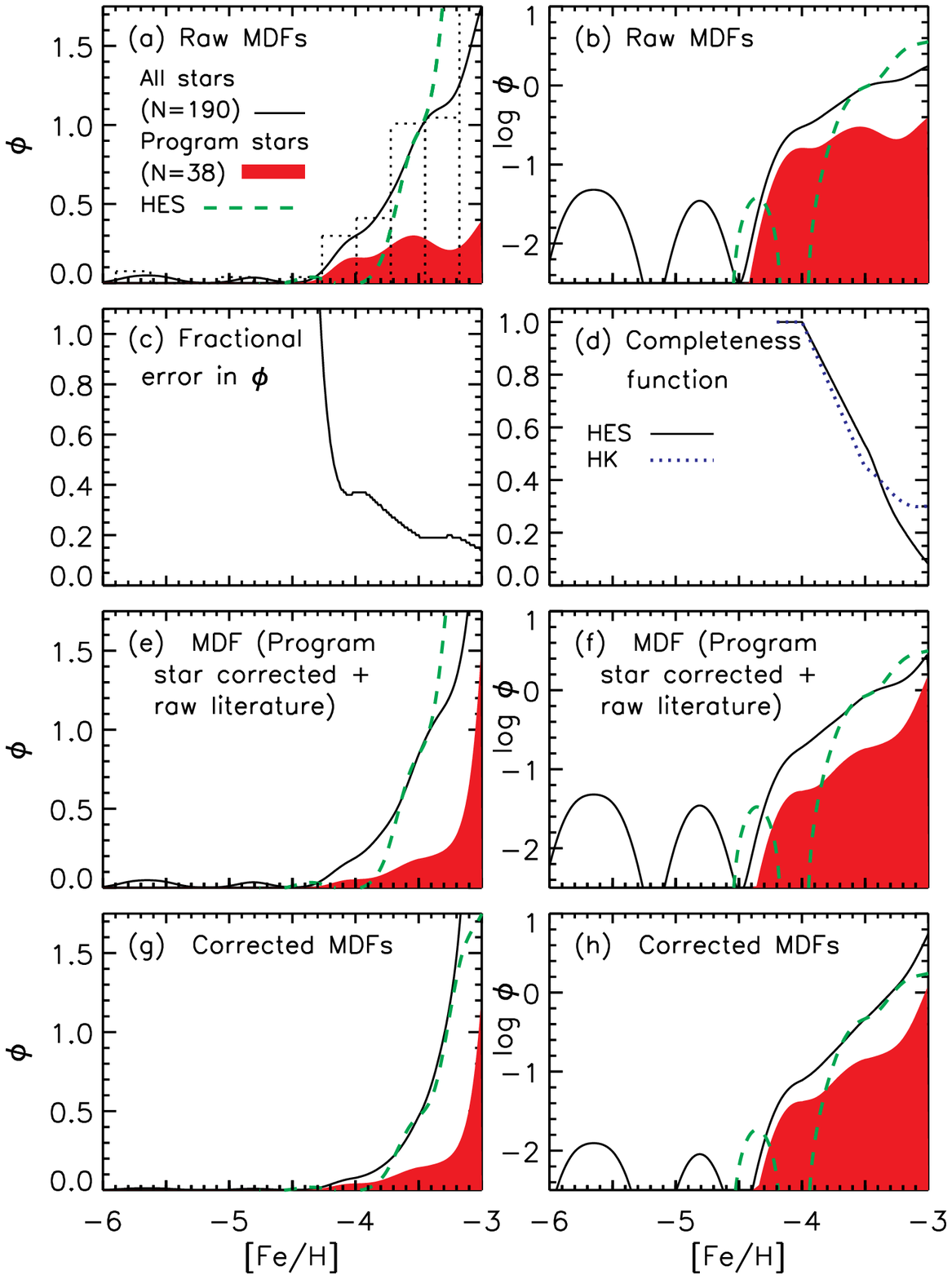} 
\caption{Generalized histograms showing the MDF (linear (left)
and logarithmic (right) scales). 
The full sample (solid black line) and program stars (red histogram)
are shown. 
The 
green dashed line is the raw HES MDF from 
\citet{schorck09}, shifted by $-$0.29 dex 
and scaled to 
match our MDF at [Fe/H] = $-$3.5. 
Panel (a) includes a 
regular histogram (dotted line) employing the \citet{shimazaki07} 
optimal bin width algorithm. 
Panel (c) shows the fractional uncertainty in our MDF (e.g, a 
fractional uncertainty of 0.2 represents an error of 20\% of the MDF value.)  
Panel (d) shows the HES and HK completeness functions. The HES 
completeness function is applied to the MDFs shown in panels (e,f,g,h). 
\label{fig:mdf}} 
\end{figure*}

Construction of our smoothed MDF includes uncertainties, which we
estimate in the following manner using Monte Carlo simulations.  We
replaced each data point, [Fe/H], with a random number drawn from a
normal distribution of width 0.15 dex, centered at the [Fe/H] of the
given data point.  We repeated this process for each data point in our
collective sample of 190 stars, and a generalized histogram was
constructed for this new sample. We repeated this process for 10,000
new random samples, producing a generalized histogram for each new
random sample. At a given [Fe/H], we then have a distribution of some
10,000 values, one for each MDF.  We measured the FWHM of this
distribution, and adopt this value as an estimate of the uncertainty
in our MDF at a given [Fe/H]. In Figure 1(c), we plot the
fractional uncertainty, where a value of 0.2 represents a 20\%
uncertainty in the value of the MDF. The relative uncertainty reaches
50\% near [Fe/H] = $-4.2$, and becomes rapidly larger at lower
metallicities, indicating that the sample size loses much statistical
significance below this value.

We also constructed a regular histogram to compare with the smoothed
MDF. We employed the \citet{shimazaki07} algorithm to determine the
optimal bin width (0.272 dex) for the full 190 star sample.  As
expected, both histograms exhibit a similar behavior.

We corrected the ``program star MDF'' using the HES
completeness function described above in Section~\ref{sec:biases}
(here shown together with the HK completeness function in Figure
1(d)), leaving the ``literature sample MDF'' unchanged. These MDFs
are presented in Figure \ref{fig:mdf} (panels e-f). We also
corrected the full MDF (i.e., ``program star + literature sample''
MDF) using the HES completeness function, and plot both corrected
MDFs in Figure \ref{fig:mdf} (panels g-h). 
While the selection biases associated with the discovery of the stars
in the SAGA database are not explicit, almost half of the 86 stars
(42) in Table \ref{tab:param} carry HK-survey names, while
most others (36) have HES-survey nomenclature. It is clear
that the majority of stars in Table 1 have been found in those
low-resolution spectroscopic surveys, and thus inherit the
spectroscopic- and volume-selection biases of those works, plus
additional biases imposed in later follow-up with medium- and
high-resolution spectroscopy. Many of the HK-survey stars would also
have been recovered in the HES survey, but were not renamed.
Consequently, using the HES completeness function should be a
reasonable step.
Given the clear similarity
between the HES and HK completeness functions below [Fe/H] = $-$3.3,
the corrected MDF would be essentially identical in this metallicity
regime had we used the HK completeness function.  

We use a two-sample Kolmogorov-Smirnov (KS) test to compare the MDFs
for dwarfs ($\log g$ > 3.5) and giants ($\log g$ < 3.5).  The null
hypothesis is that the dwarf and giant MDFs are drawn from the same
distribution.  For [Fe/H] $\le$ $-$3.0, the two-sample KS test yields
a probability of 0.601 (D = 0.167) that the dwarf and giant MDFs are
drawn from the same distribution\footnote{The dwarf and giant MDFs for
[Fe/H] $\le$ $-$3.0 may be seen in Figure \ref{fig:cdf}
panels (d) and (e), respectively, which we shall discuss in what
follows.}.  A similar test for [Fe/H] $\le$ $-$3.5 yields a
probability of 0.915 (D = 0.200) that the dwarf and giant MDFs are
drawn from the same distribution.  Therefore, the null hypothesis that
the giants and dwarfs are drawn from the same population cannot be
rejected at the 0.10 level of significance, the least stringent level
in Table M of \citet{siegel56}.

In Figure \ref{fig:mdf}(a), we overplot the raw MDF from \citet{schorck09}
(using the values in their Table 3).  Comparing our sample with
\citet{schorck09} and \citet{li10} (made available by N.\ C.),
we find 12 stars in common.  For these 12 stars, there are some 18
[Fe/H] measurements that can be compared.  For the nine program stars 
for which we conducted dwarf and subgiant analyses, 
we treat both [Fe/H] values as
independent measurements for the purposes of this comparison.  Our
metallicities differ from the \citet{schorck09} and \citet{li10} 
values by $-$0.26 $\pm$ 0.06
dex ($\sigma$ = 0.27 dex), and so we shift the raw HES MDF 
of \citet{schorck09} by $-$0.26
dex in Figure \ref{fig:mdf}, and scale it to match our
MDF at [Fe/H] = $-$3.5.  Below [Fe/H] = $-$3.5, we find a large
fraction of stars relative to the HES MDF. 
In Figure \ref{fig:mdf}(f), both the program star and literature sample MDFs 
have a slope close to 1.0, consistent with the Hartwick Simple Model, 
down to the shoulder at [Fe/H] $\simeq$ $-4.1$, when the finite sample 
begins to run out of stars (which are necessarily counted in integers). 
This corresponds to the metallicity at which the fractional error 
(Figure \ref{fig:mdf}(c)) increases rapidly and, as noted above, the 
finite sample size loses much statistical significance. 

Therefore, taken at face value, and bearing in mind the 
biases, the apparent cutoff in the HES MDF at [Fe/H] = $-$3.6 is not
confirmed by our data.  We identify 13 HES stars in our sample that
have [Fe/H] $\le$ $-$3.7 (of which four are contained in the work of
\citealt{schorck09} and \citealt{li10}).  We speculate that (i)
stars in our sample having [Fe/H] $<$ --3.7 were
rejected as having strong $G$-bands 
(GP\footnote{This is the \citet{beers99} index that measures the
strength of the 4300\AA\ CH molecular features.} > 6 \AA), and/or (ii) our
abundance scale differs from that adopted in the
\citet{schorck09} and \citet{li10} analyses.

Regarding point (i), none of our objects has
GP > 6\AA.  In
particular, we note that the three most Fe-poor HES stars, all of
which have large [C/Fe] ratios, are not rejected by this criterion.
Concerning point (ii), Figure~\ref{fig:fe_err} shows the metallicity
difference $\Delta$[Fe/H] = [Fe/H] (high resolution: this study) $-$
[Fe/H]$_{\rm K}$ (medium resolution: \citealt{schorck09,li10}) versus
\teff, $\log g$, [Fe/H], E($B-V$), and GP.  In each panel, we plot the
linear least squares fit to the data, and show the formal slope and
uncertainty as well as the dispersion about the slope.  In all cases,
the dispersion about the slope is compatible with the value expected
based on the convolution of the errors, $\sigma$ (combined) = 0.25 dex
assuming $\sigma$ (this study) = 0.15 dex and $\sigma$ \citep{beers99}
= 0.20 dex.  The correlation between $\Delta$[Fe/H] versus [Fe/H] is
significant at the 2-$\sigma$ level (although we caution that the errors 
on these two quantities are correlated), indicating that as one moves to
lower metallicity, the [Fe/H] values from our high-resolution analysis
are lower than those based on medium-resolution spectra.  Such a
correlation would help, in part, to explain why we do not find a
cutoff in the MDF.  Possible explanations for this correlation include
systematic differences in the analyses, interstellar Ca absorption
lines, and/or CH molecular stellar lines in the region of the Ca\,{\sc
ii} K line.  Further insight into the 
differences from high-resolution and medium-resolution spectra await
larger comparison samples.  
(We also note that Figure \ref{fig:fe_err} 
includes 3-$\sigma$ correlations between $\Delta$[Fe/H] and \teff\ (panel a)
and $\Delta$[Fe/H] and \logg\ (panel b).) 
For completeness, we note that linear
regression analysis shows that the best fit to $\Delta$[Fe/H] (high
resolution $-$ medium resolution) is $-$1.825 +
2.442$\times10^{-4}\times$\teff\ + 0.151$\times \log g$
+0.160$\times$[Fe/H]$_{\rm high~resolution}$ $-$ 0.401$\times$E($B-V$) 
+ 0.337$\times$GP.  The dispersion about this fit is 0.17 dex, and 
the uncertainties in the coefficients are 
2.986$\times10^{-4}$, 0.147, 0.189, 6.637, and 0.295 for 
\teff, \logg, [Fe/H], E$(B-V)$, and GP, respectively. 

\begin{figure*}[t!]
\epsscale{0.6}
\plotone{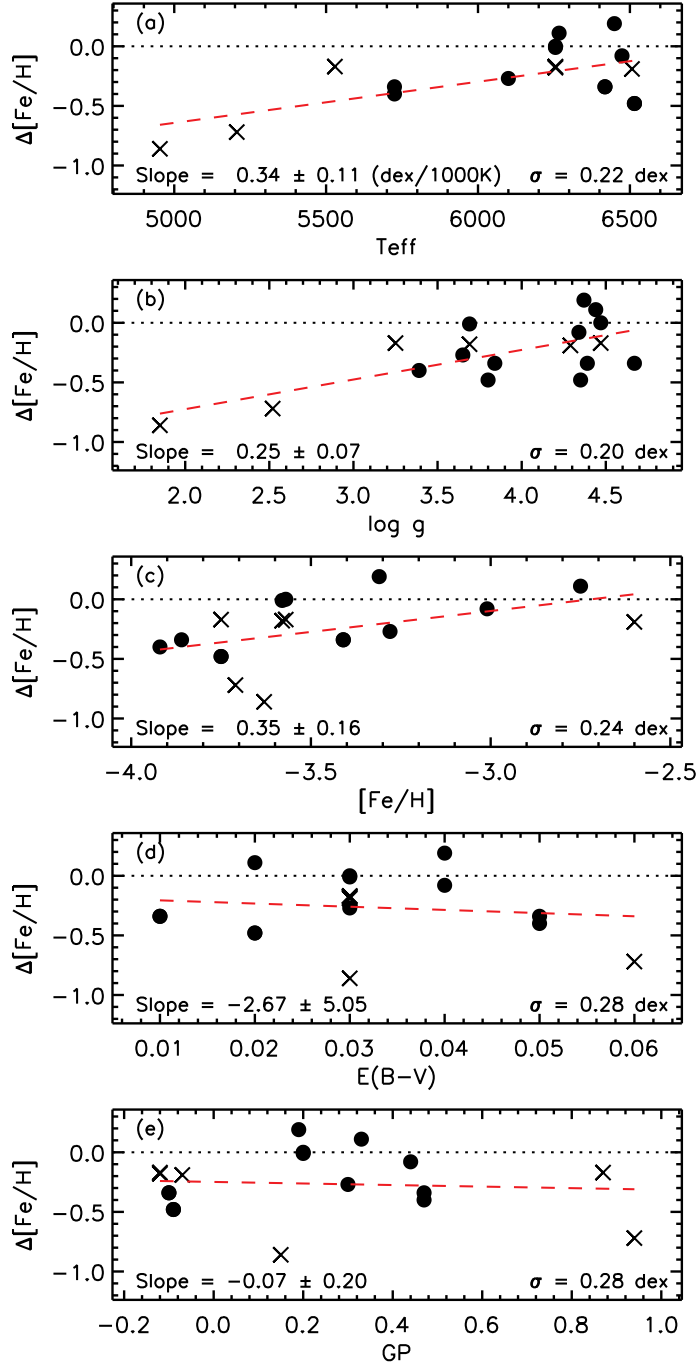} 
\caption{The difference in metallicity (This Study $-$ Literature) 
between our analysis and those
of \citet{schorck09} (crosses) and \citet{li10} (circles) vs.\ (a)
\teff, (b) $\log g$, (c) [Fe/H], (d) E($B-V$), and (e) GP, 
where the abscissa values in panels (a), (b), and (c) 
were obtained from the high-resolution analysis. In each
panel, we plot the linear least squares fit to the data, and show the
slope, uncertainty, and dispersion about the slope.  In this figure,
we include both the dwarf and subgiant [Fe/H] measurements for those
program stars with multiple analyses (see
Section~\ref{sec:obs} for details).  \label{fig:fe_err}
}
\end{figure*}

In Figure \ref{fig:mdfcomp}, we compare the raw and corrected MDFs
with several model predictions, scaled to match our MDFs at
[Fe/H] = $-$3.5.  The rationale for choosing this
normalization is that (i) in this metallicity regime we
expect that our sample includes the vast majority of stars currently
known, albeit with selection biases, and (ii) we hope to provide a
more detailed consideration of the MDF at the lowest
observed [Fe/H] values.

\begin{figure*}[t!]
\epsscale{0.8}
\plotone{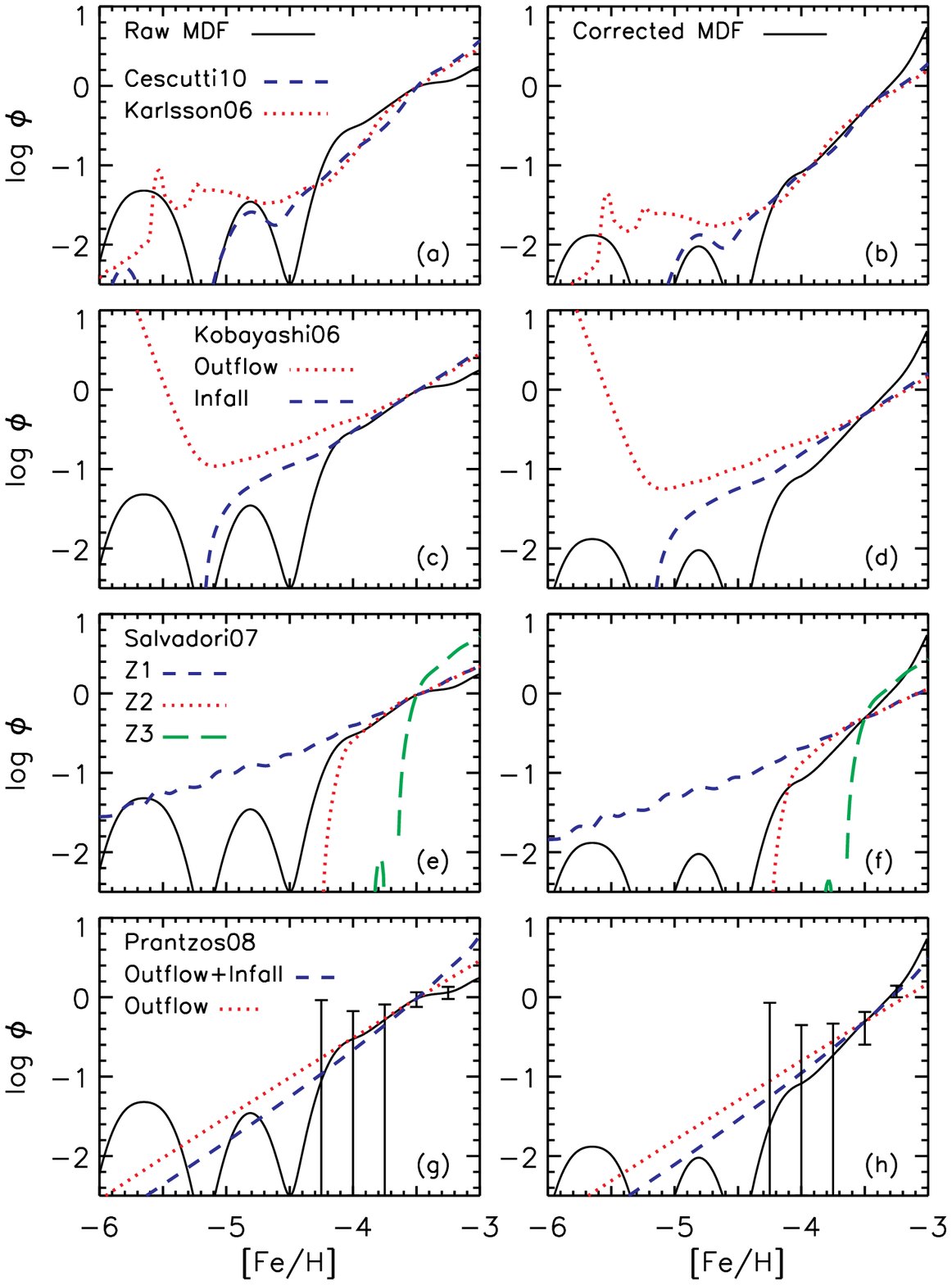} 
\caption{Comparison of the raw (left) and corrected (right) MDFs with the 
\citet{karlsson06}, \citet{kobayashi06}, 
\citet{salvadori07}, \citet{prantzos08}, and 
\citet{cescutti10} predictions. 
The predictions are scaled to match our MDF at [Fe/H] = $-$3.5. 
In panels (e,f), the Z1, Z2, and Z3 lines represent critical metallicities of 
Z$_{\rm cr}$ = 0, 10$^{-4}Z_\odot$, and 10$^{-3.4}Z_\odot$ 
respectively. 
In the lower panels, we show error bars on our raw and corrected MDFs. 
\label{fig:mdfcomp}}
\end{figure*}

All predictions, except the \citet{kobayashi06} ``outflow'' model, 
provide a reasonable fit to the raw and
corrected MDFs.  The \citet{kobayashi06} ``infall'' model provides a
superior fit to our MDF than their ``outflow'' model (which overpredicts
the number of metal-poor stars).  
The ``outflow'' model contains (i) outflow, (ii) no infall, and 
(iii) a low star-formation efficiency, while the ``infall'' model contains
(i) no outflow, (ii) infall, and (iii) a much lower star-formation 
efficiency.  
\citet{prantzos08} adopts a
hierarchical merging framework in which the halo is formed from
sub-halos, with a distribution in stellar mass, and with the MDF of each
sub-halo based on Local Group dwarf satellite galaxies. Both Prantzos models 
(``outflow'' only and ``outflow+infall'') provide equally good fits to
our MDF.  \citet{salvadori07} provide predictions  
for different
critical metallicities, $Z_{\rm cr}$, and their $Z_{\rm cr}$ =
10$^{-4}Z_\odot$ and $Z_{\rm cr}$ = 0 models both provide reasonable
fits to our MDF.  The raw and corrected MDFs indicate that the
critical metallicity, above which low-mass star formation is possible,
is well below $Z_{\rm cr}$ = 10$^{-3.4}Z_\odot$, in contrast to the
\citet{schorck09} and \citet{li10} MDFs.  
In addition to the spectroscopic selection biases noted earlier, 
we need to be mindful of possible volume-selection biases, and that 
the real Galactic MDF at low metallicities could be significantly 
different from the one presented in this paper. Still larger, 
deeper samples, the biases and completeness of which are better understood, 
are necessary to obtain this MDF. 

\subsubsection{On the nature of the MDF}

We now explore four aspects of our MDF analysis: 
(1) choice of a lower-metallicity cutoff versus a higher-metallicity cutoff, 
(2) usage of a regular histogram versus a generalized histogram, 
(3) adoption of a linear versus a logarithmic scale, and 
(4) inclusion of elements in addition to Fe when defining the metallicity. 

~

{\it Lower-metallicity cutoff versus higher-metallicity cutoff} 

~

In order to explore the first aspect, we adopt the 
(one-zone, closed-box) Simple Model of Galactic
chemical evolution \citep{schmidt63,searle72,pagel75,hartwick76}, and
create two MDFs, from which we remove all stars below [Fe/H] = --4.5
(lower-metallicity cutoff) and --4.0 (higher-metallicity cutoff,
sometimes referred to as a ``sharp cutoff''). Both are populated with
stars on a regular grid of step size 0.05 dex, and normalized such
that they have 1000 stars below [Fe/H] = $-$3.0, i.e.,
some 12 times larger than our 86 star sample in Table
\ref{tab:param}.  For the lower-metallicity cutoff (MDF1), there are
four stars in the lowest metallicity bin, [Fe/H] = $-$4.5, while for
the higher-metallicity cutoff (MDF2), there are 20 stars in the lowest
metallicity bin, [Fe/H] = $-$4.0.  The two MDFs are shown in Figure
\ref{fig:mdf2}. In the upper panel, one sees that when overplotted on the full
metallicity range, $-$5.0 $<$ [Fe/H] $<$ 0.0, they are indistinguishable. 
When considering only the regime below [Fe/H] = $-$3.0
(Figure 4 panels (b) and (c)), however, the difference in
the two MDFs is clear.

\begin{figure*}
\epsscale{0.8}
\plotone{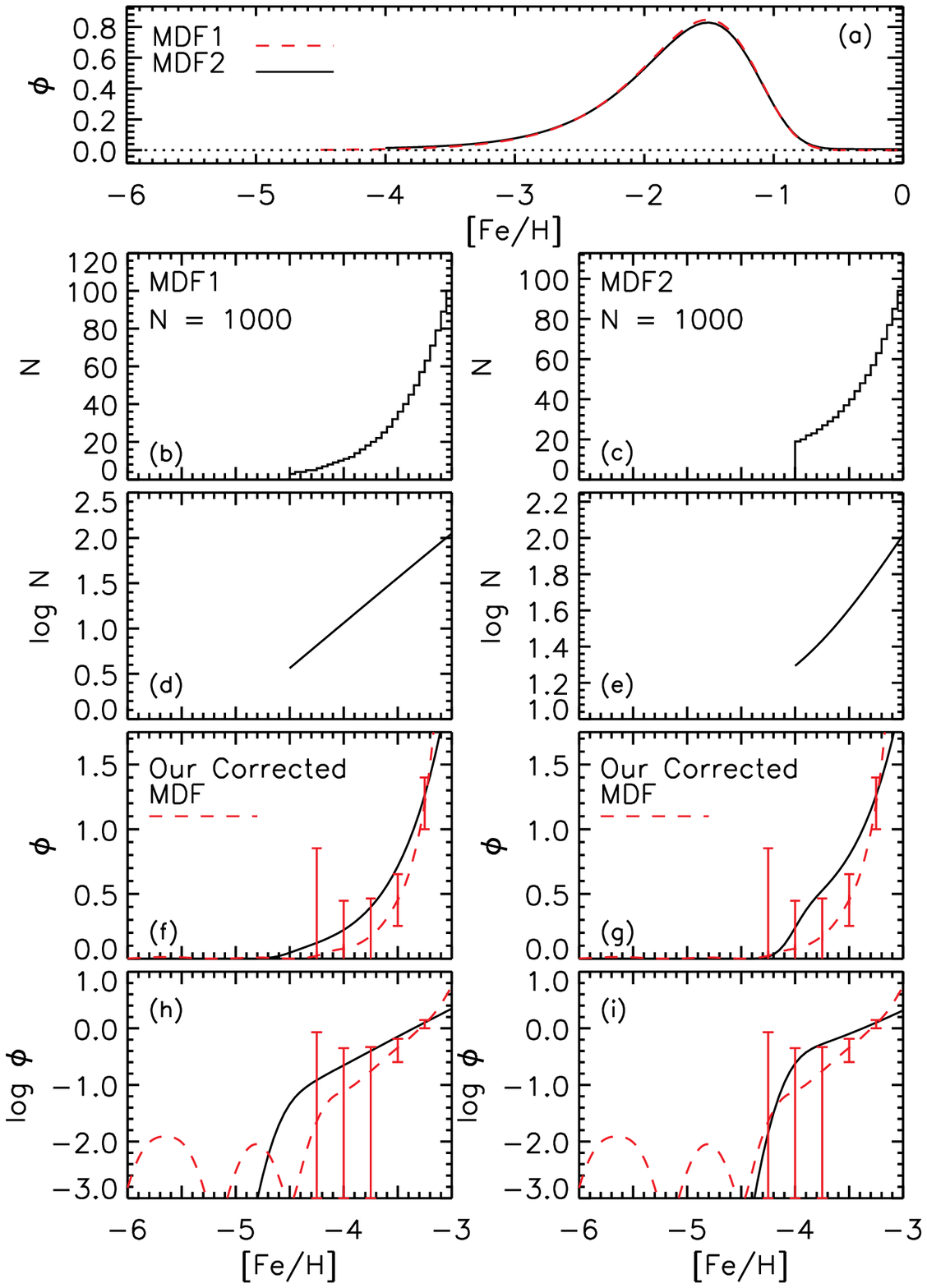} 
\caption{
Comparison of the lower-metallicity cutoff (MDF1) and higher-metallicity 
cutoff (MDF2) cases 
generated using the Simple Model.  Panels (a,b,c,f,g) are on a linear
scale while panels (d,e,h,i) are on a logarithmic scale. Panels (b)
through (e) are regular histograms while panels (f) through (i) are
generalized histograms. Our corrected MDF is overplotted in panels (f)
through (i) along with error bars. 
Panel (f) is normalized so the total area is 1.0, and panel (h) is 
produced directly from this panel. In panels (f) through (i), we 
overplot our corrected MDF adopting the error bars from the raw
MDF (we do not show error bars below [Fe/H] = $-4.25$ since they 
extend beyond the range of these plots). 
\label{fig:mdf2}
}
\end{figure*}

~

{\it Regular histogram versus generalized histogram}

~

Panels (b) and (c) of Figure \ref{fig:mdf2} show regular histograms
for the two MDFs, while panels (f) and (g) show generalized
histograms.  As expected, the generalized histogram smooths out 
the data along the abscissa. Given the 
numbers of stars in the lowest metallicity bins, the 
lower-metallicity cutoff 
MDF may appear to have an ``extended tail,'' when 
represented in generalized histogram format, but in reality, 
both MDFs now have an additional tail. 

~

{\it Linear versus logarithmic scale}

~

Panels (b,c,f,g) and (d,e,h,i) of Figure \ref{fig:mdf2} have 
linear and logarithmic scales respectively. Panels (d) and (e) 
(regular histograms) and panels (h) and (i) (generalized histograms) 
exhibit rather similar trends. When using a logarithmic scale, it is 
easier to discern where the MDF cuts off, (as every finite sample, 
observed or simulated, must). The generalized histogram replaces 
each datum with a Gaussian function, and taking the logarithm of 
this yields an inverted quadratic function; i.e., each datum 
contributes an inverted quadratic function to the log panel. 
In Figure \ref{fig:mdf2}(h), the last Monte Carlo datum at 
[Fe/H] = $-4.5$ gives rise to the quadratic roll-off at 
[Fe/H] $<$ $-4.5$, and in Figure \ref{fig:mdf2}(i) the last 
Monte Carlo datum at [Fe/H] = $-4.0$ gives rise to the roll-off 
at [Fe/H] $<$ $-4.0$. This roll-off meets the populated part of 
the MDF at a ``shoulder'', above which the MDF rises with a 
slope of 1.0, due to the adoption of the Simple Model. The location 
of the shoulder indicates the metallicity at which either the finite 
sample size becomes too small to populate the MDF, as in this 
simulation, or the MDF genuinely departs from the Galactic 
chemical evolution model pertaining at higher metallicity, 
as would be the case in the scenarios envisaged by \citet{salvadori07} 
and others discussed in connection with Figure \ref{fig:mdfcomp}. 
The fact that the shoulder in our observed MDF 
(e.g., Figures \ref{fig:mdf}(f), 1(h) or Figure \ref{fig:mdfcomp}), 
determined from high-resolution spectroscopic analyses, 
is located at [Fe/H] = $-4.1$ or $-4.2$, and attains a slope close to 
1.0 at higher metallicity, gives us the confidence that the MDF does 
not exhibit a sharp drop at [Fe/H] = $-3.6$, nor indeed in the 
metallicity range down to [Fe/H] = $-4.1$.

~

{\it Inclusion of elements in addition to Fe in the ``metallicity''}

~

Strictly defined, metallicity ($Z$) includes all elements heavier than
helium, although in practice Fe is widely adopted as the canonical
measure of stellar metallicity. Therefore, the MDF discussed thus far is
really the Fe distribution function.  For the Sun, the seven most
abundant metals, in decreasing order, are O, C, Ne, N, Mg, Si, and Fe
\citep{asplund09}.  Therefore, in order to explore this
fourth aspect of our discussion, the behavior of the MDF when
including additional elements, we arbitrarily define $Z$ to consist of
C, N, Mg, Si, and Fe.  (Of the 86 stars with [Fe/H] $\le$ $-$3.0,
there are measurements of C, N, Mg, and Si for 54, 36,
81, and 36 stars respectively.)  We compute $Z$ for each star, only
considering the set of elements with measurements; that is, we ignore
those elements not measured in a given star.  In Figure
\ref{fig:mdfz}a we plot [Z/H] versus [Fe/H], including all stars in our
sample (N = 190).  Panels (b) and (c) in Figure \ref{fig:mdfz} show
the MDFs for [Fe/H] and [Z/H], respectively, in regular and
generalized histogram format.  The regular histogram again uses the
\citet{shimazaki07} optimal bin width algorithm (0.278 dex for [Z/H]).
We note that the two MDFs exhibit a similar behavior. 
Indeed, the [Fe/H] and [Z/H] MDFs have almost identical 
average gradients over the plotted range.  
The construction of the [Z/H] MDF, based on large samples of
stars having O and C measurements, would be of 
great interest given the postulated importance of these elements 
for low-mass star formation in the early Universe
\citep{bromm04,frebel07b}. 
Additionally, when considering the [Z/H] MDF, 
we need to be mindful of issues including 
(a) giants, in general, offer a larger suite of measurable elements than dwarfs, 
(b) for a fixed abundance, the lines in giants are generally stronger than in dwarfs, 
thereby enabling measurements in giants, rather than limits for dwarfs, in many cases, 
and (c) the highest values of $Z$ in Figure \ref{fig:mdfz}c likely suffer from 
large incompleteness. Furthermore, we note that inclusion of C, N, and O 
abundances may considerably alter the [Z/H] MDF compared to our present distribution. 
(We emphasize again that throughout the
present paper the MDF refers to the [Fe/H] distribution function
unless specified otherwise.) 

\begin{figure*}
\epsscale{0.8}
\plotone{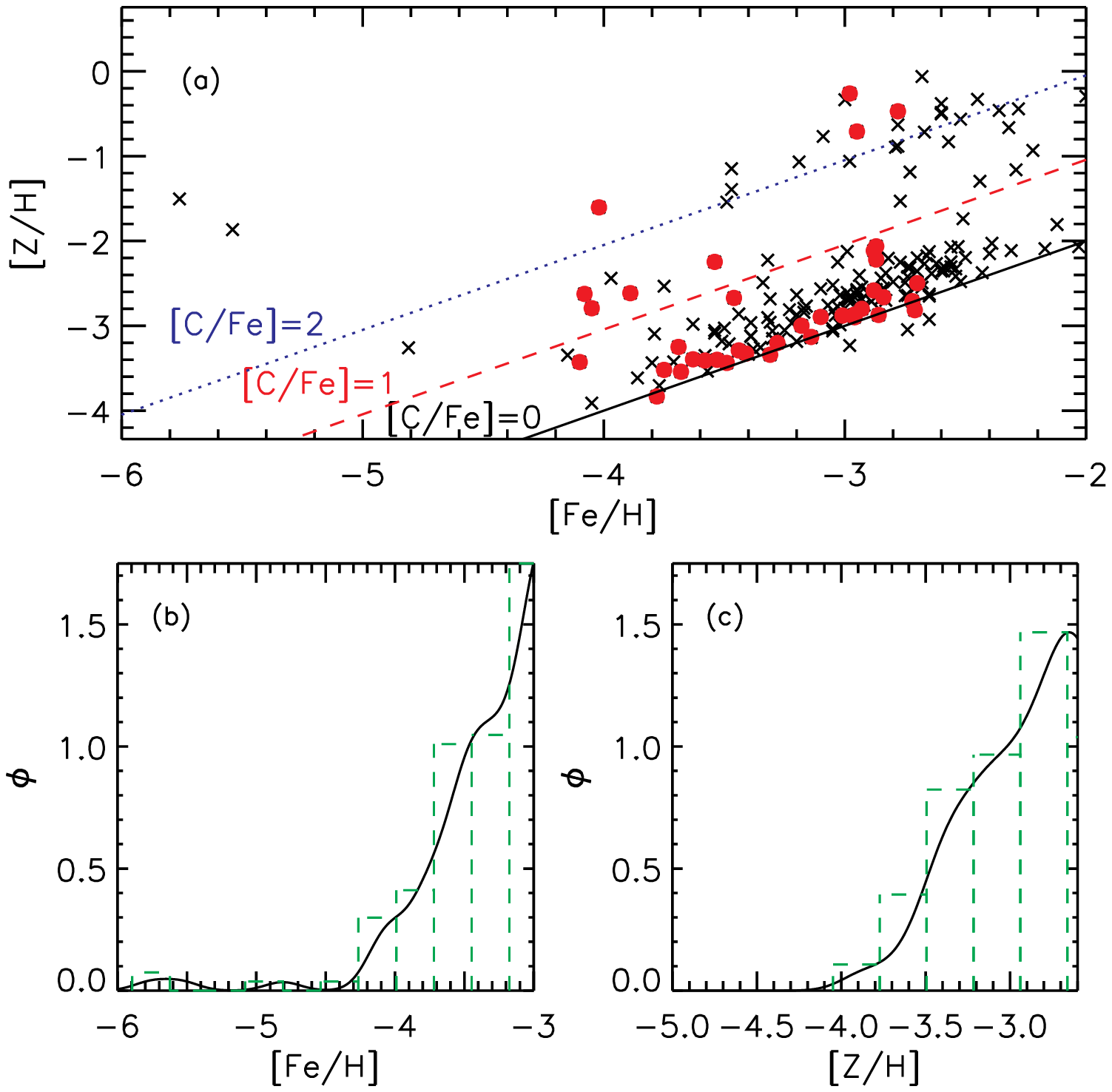} 
\caption{
[Z/H] vs.\ [Fe/H] for the full sample of stars (N = 190). 
Program stars are plotted as red circles. 
Panel (a) includes lines of constant [C/Fe]. 
In panels (b) and (c), we show regular and generalized histograms 
for [Fe/H] and [Z/H], respectively, where Z includes the available 
set of C, N, Mg, Si, and Fe abundances in a given star.  
\label{fig:mdfz}
}
\end{figure*}

Armed with sufficient observational data, MDFs can of course be 
constructed for a range of specific elements, e.g., [O/H] and [C/H], 
rather than just for [Fe/H] or [Z/H]. Such element-specific MDFs 
can then be compared with the outputs of various chemical evolution 
models, as we did for [Fe/H] in Figure \ref{fig:mdfcomp}. 
Doing so may provide valuable insights into the triumphs and 
deficiencies of those models, and indicate ways in which they 
can be improved.

\subsection{The Fraction of Carbon-Enhanced Metal-Poor (CEMP) Stars}

In Figure \ref{fig:cdf}, we again plot the raw MDF (using generalized
histograms), but on this occasion we also include in the figure the
MDF when restricted to CEMP objects, where we have used the CEMP definition of
\citet{aoki07} 
([C/Fe] $\ge$ +0.70, for $\log (L/L_\odot) \le 2.3$ and [C/Fe] $\ge$ 
+3.0 $-$ $\log (L/L_\odot)$, for $\log (L/L_\odot) > 2.3$; 
as opposed to the \citet{beers05} definition of [C/Fe] $>$ +1.0).  
In panel (c) we show
the percentage of CEMP stars as a function of [Fe/H], which we obtain
by dividing the CEMP MDF by the MDF containing only those stars with 
C-measurements or C-limits below the CEMP threshold. 
(Here we present results using the CEMP
definitions of both \citealt{aoki07} and \citealt{beers05}.) 
Using Monte Carlo
simulations, as described earlier, we estimate the fractional
uncertainty in the CEMP MDF, and therefore the uncertainty in the CEMP
percentage at a given [Fe/H].    Note that
for our 38 program stars from Paper I, C abundances (or limits) were
measured from the spectra. For the literature sample, we were unable
to conduct the necessary spectrum synthesis re-analysis (since we did
not have access to the spectra), and we chose not to make any
adjustments to these abundances based on our adopted stellar
parameters and metallicity\footnote{Had we updated the [C/Fe] ratio 
via [C/Fe]$_{\rm New}$ = [C/Fe]$_{\rm Literature}$ $-$ ([Fe/H]$_{\rm
This~study}$ $-$ [Fe/H]$_{\rm Literature}$), the numbers of CEMP
objects would change from 16 to 19 and from 22 to 28 for the
\citet{beers05} and \citet{aoki07} definitions, respectively. 
However, we note 
that this approach only includes changes to the metallicity, and does 
not address any changes in the C abundance.}.  We
also note that for stars with large [C/Fe] ratios (and for metal-poor
stars in general), a more rigorous chemical abundance analysis would
require, amongst other things, model atmospheres with appropriate CNO
abundances and consideration of 3D and/or non-LTE effects \citep{asplund05}. 
Bearing in mind these shortcomings, as well as issues
regarding selection biases and completeness of our sample already
discussed, we now comment on the fraction of CEMP stars.

\begin{figure*}
\epsscale{0.8}
\plotone{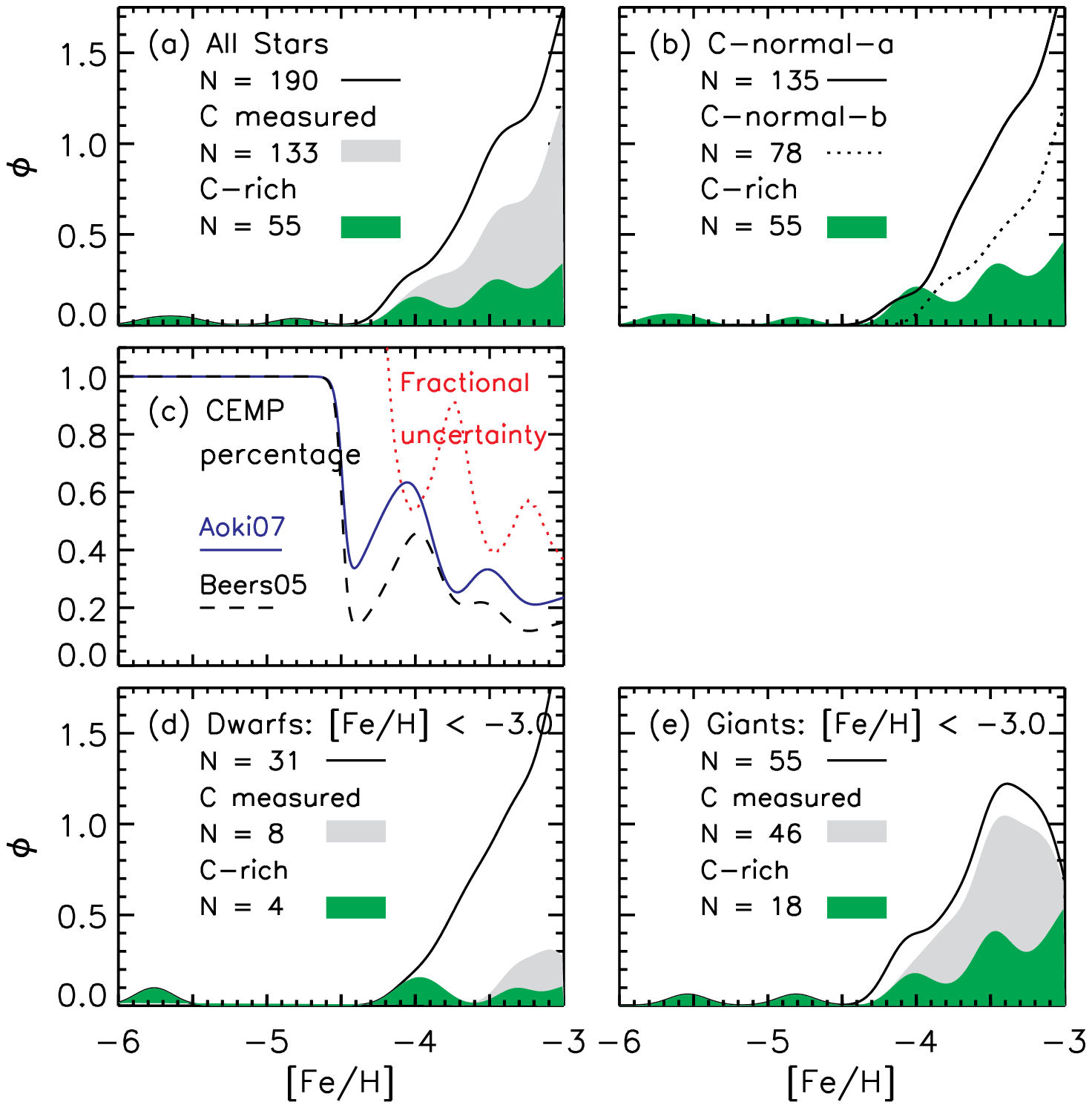} 
\caption{
Generalized histograms showing the raw MDF for 
all stars (solid black line), CEMP objects (green histogram), 
and stars for which [C/Fe] was measured (grey histogram). 
Panel (b) shows the MDF for (i) the C-normal population including 
stars with [C/Fe] limits (C-normal-a), (ii) the C-normal population excluding 
stars with [C/Fe] limits (C-normal-b), 
and (iii) the CEMP sample. 
Panel (c) shows the CEMP fraction and the fractional uncertainty. 
Panels (d) and (e) show dwarfs and giants, respectively.  
\label{fig:cdf}
}
\end{figure*}

We find a CEMP fraction of 32 $\pm$ 8\% (22 of 69) adopting the
\citet{aoki07} criteria\footnote{If we had considered
stars with [Fe/H] $\le$ $-$2.80, an arbitrarily chosen more
metal-rich boundary, we would have obtained a CEMP fraction of 29
$\pm$ 6 \% (28 of 98), using the \citealt{aoki07} definition.},
and 23 $\pm$ 6\% (16 of 71) using the \citet{beers05} criterion for
[Fe/H] $\le$ $-$3.0.  (As noted above, in determining the CEMP
fraction we only adopt stars for which we had C measurements or C
limits below the CEMP threshold. Thus, the total number of stars is
not 86.  Adopting the \citet{aoki07} criteria, we find CEMP
fractions of 25 $\pm$ 8\% (11 of 44) and 29 $\pm$ 15\% (5 of 17) in
the metallicity ranges $-3.5$ < [Fe/H] $\le$ $-$3.0 and $-4.0$ <
[Fe/H] $\le$ $-$3.5, respectively.  Previous estimates of the CEMP
fraction below [Fe/H] = $-$2.0, using the \citet{beers05} [C/Fe] $\ge$
+1.0 criterion, include 14 $\pm$ 4\% \citet{cohen05}, 9 $\pm$ 2\%
\citet{frebel06}, and 21 $\pm$ 2\% \citet{lucatello06}, all of which
are probably comparable with our value, given the differences in
[Fe/H] ranges for the samples.  For the 38 program stars of Paper I,
there was a bias towards CEMP objects. Our somewhat subjective
observing criteria at the Keck and Magellan telescopes, as applied to
an evolving candidate list, was to (i) observe the most metal-poor
candidates available, (ii) in the event of similar metallicity
estimates, prefer giants over dwarfs, and (iii) for more metal-rich
candidates, observe objects with prominent $G$-bands in their
medium-resolution spectra, with the expectation that a small fraction
might be C-rich, $r$-process enhanced stars similar to CS 22892-052,
some of which might have measurable Th and U for cosmo-chronometric
age determinations (e.g., \citealt{barklem05,sneden08}).

Within our sample, the CEMP fraction is higher for dwarfs 
(50 $\pm$ 31\%; 4 of 8) than for giants (39 $\pm$ 11\%; 18 
of 46).  This discrepancy may reflect the fact that, for a fixed
metallicity and [C/Fe] abundance ratio, the CH molecular lines are
stronger, and therefore more likely to yield a measurement, in giants
than in dwarfs. That is, some of our dwarfs have such high [C/Fe]
limits that they may indeed have [C/Fe] $\ge$ +0.7, and thus the CEMP
fraction for dwarf stars is very likely an upper limit.  Indeed, some 23
of 31 (74 $\pm$ 20\%) dwarf stars have C limits (or no
measurements), compared with only 9 of 55 (16 $\pm$ 6\%) giant
stars. 

There are previous reports in the literature that the CEMP fraction rises with
decreasing metallicity (see \citealt{carollo12} for a full description). 
Including the \citet{caffau11} object, three
of the four stars with [Fe/H] $\le$ $-$4.5 are CEMP objects.  For our
sample, of the 65 stars with $-$4.3 $\le$ [Fe/H] $\le$ $-$3.0 and [C/Fe] 
measurements, 19 are
CEMP objects. Adopting this CEMP fraction of 0.29, the
probability of having three CEMP objects in a sample of four stars, as
is the case for [Fe/H] $\le$ $-$4.5, is
0.076. While further data are clearly necessary to
settle the issue, relative carbon richness at the lowest values of
[Fe/H] seems ubiquitous. 
We refer the reader to \citet[and references therein]{carollo12}, who
demonstrate that the CEMP fraction increases from 0.05 to 0.26 $\pm0.03$ as
[Fe/H] as the metallicity decreases from [Fe/H] = $-$1.5 to [Fe/H] = $-$2.8, based on
a large sample of calibration stars from the Sloan Digital Sky Survey
(SDSS; \citealt{york00,gunn06}). 

The above comments notwithstanding, in Figure
\ref{fig:cdf2} we plot the CEMP fraction as a function of [Fe/H]
(upper panel) and [Z/H] (lower panel).  For $-$4.5 $\le$ [Fe/H]
$\le$ $-$3.0, we have three bins with roughly equal numbers, while
the fourth bin, [Fe/H] $\le$ $-$4.5, has only 3 stars.  
There is no significant correlation between the CEMP fraction 
in each bin at the median [Fe/H] of each bin; Figure \ref{fig:cdf2}(a) 
suggests a slope of $-$0.24 $\pm$ 0.22. 
Had we included the C-normal ultra metal-poor
\citet{caffau11} star, we would have obtained a slope of 
$-$0.20 $\pm$ 0.19. 
For [Z/H], we use
four bins with equal numbers of stars.  We again measure the linear
fit to the CEMP fraction at the median [Z/H] of each bin.  
In this case, there is no significant correlation between the 
CEMP fraction in each bin at the median [Z/H] of each bin; 
Figure \ref{fig:cdf2}(b) suggests a slope of 0.03 $\pm$ 0.10. 
An important consideration is that the sample 
was selected to have low metallicity such that the stars with 
the highest $Z$ tend to have high C abundances. Such a bias may 
potentially explain the positive trend we find between CEMP fraction and [Z/H]. 
Thus, we reiterate the need to measure O and N when possible to 
better define the metallicity, $Z$. 
Nevertheless, we caution that the behavior of the CEMP fraction at
lowest metallicity likely depends on the adopted ``metallicity''
definition. 

\begin{figure*}
\epsscale{0.8}
\plotone{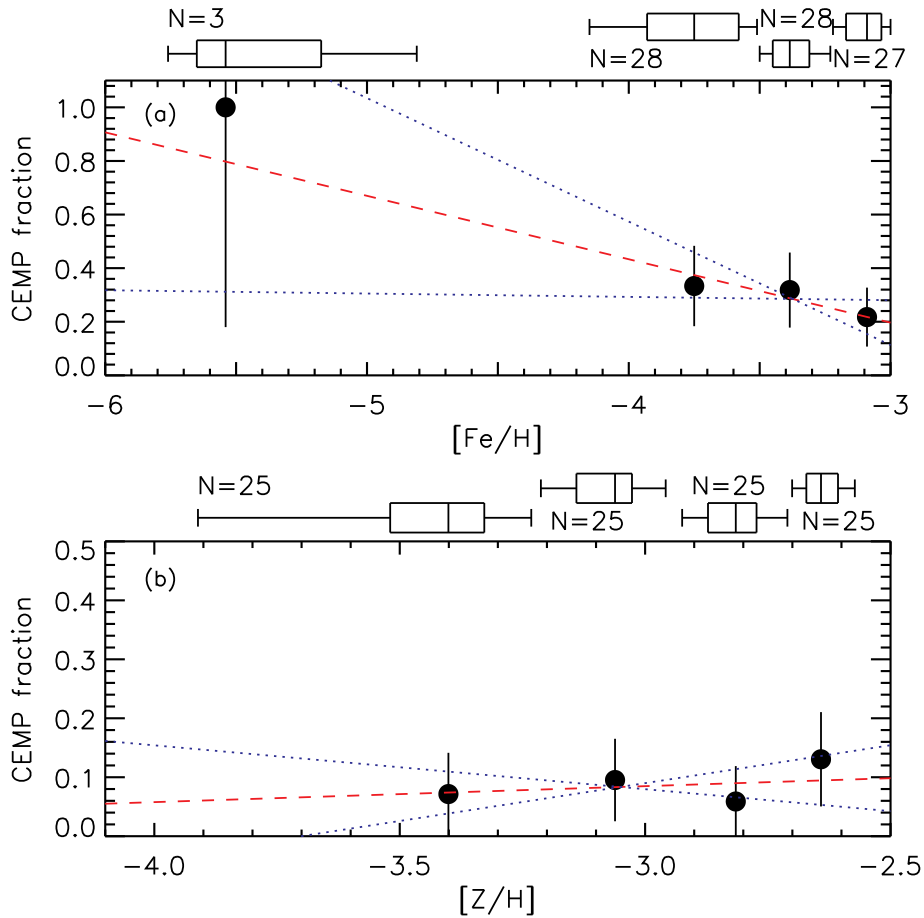} 
\caption{CEMP fraction versus [Fe/H] (upper) and [Z/H]
(lower). In the upper panel, the lowest metallicity bin covers
[Fe/H] $\le$ $-$4.5 while the three remaining metallicity bins
have roughly equal numbers of stars.  In the lower panel, the four
metallicity bins contain equal numbers of stars.  The boxplots
above both panels show the distributions in metallicity and the
numbers of stars per bin.  In both panels, the red dashed line is
the linear fit to the binned data (slope and uncertainty are 
given) and the blue dotted line shows
the 1-$\sigma$ uncertainties to the best fit. 
\label{fig:cdf2}
\vspace{8mm} 
}
\end{figure*}

\section{CONCLUDING REMARKS}

We have conducted a homogeneous abundance analysis of extremely
metal-poor stars from an equivalent-width analysis based
on high-resolution, high-S/N spectra.  Our sample contains 86  
objects with [Fe/H] $\le$ $-$3.0, including 32 below [Fe/H] = $-$3.5.
While the completeness functions for our $\sim$ 30 HES
program stars and the $\sim$ 50 HK stars in the extended sample are
well understood, the selection biases for the remaining literature
sample are poorly known.  Nevertheless, our results provide an
important new view of the MDF and CEMP fraction at lowest metallicity.

The raw and corrected MDFs do not show evidence for a cutoff at [Fe/H]
= $-$3.6. Both MDFs appear to decrease smoothly down to 
at least [Fe/H] = $-$4.1. 
Four stars with much lower metallicity are also known, 
three of which are present in our sample 
(the fourth being SDSS J102915+172927; \citealt{caffau11}). 

The fraction of CEMP stars in our sample below [Fe/H] = $-$3.0 is
23 $\pm$ 6\% and 32 $\pm$ 8\%, when adopting the
\citet{beers05} and \citet{aoki07} definitions, respectively.  The
former value is in good agreement with previous estimates (based on 
the \citealt{beers05} criterion).  
It is unclear whether the CEMP fraction increases with decreasing 
metallicity below [Fe/H] = $-3$, as the apparent trend is not statistically 
significant ($<$ 1$\sigma$) in the present dataset. 

This study has pushed the boundary for any possible cutoff of the 
MDF down to at least [Fe/H] $<$ $-4.1$, but stars below this 
metallicity are already known. Exploring the regime below [Fe/H] = $-4$ requires 
still larger samples of metal-poor stars, coupled with a more
rigorous analysis that includes non-LTE effects, 3D hydrodynamical
model atmospheres and, due to the prevalence of CEMP stars, 
appropriate CNO abundances.  Upcoming surveys
will hopefully produce significant numbers of metal-poor stars in the
near future to address this need.

\acknowledgments

We thank C.\ Chiappini, T.\ Karlsson, C.\ Kobayashi,
N.\ Prantzos, and S.\ Salvadori for sending electronic data, and 
A.\ Alves-Brito, A.\ Karakas, and 
R.\ Izzard for helpful discussions. 
We thank the anonymous referee for a careful reading 
of the paper and helpful comments. 
D.\ Y., J.\ E.\ N., M.\ S.\ B., and M.\ A.\ gratefully acknowledge
support from the Australian Research Council (grants DP03042613,
DP0663562, DP0984924, and FL110100012) for studies of the Galaxy's most
metal-poor stars and ultra-faint satellite systems.  
J.\ E.\ N.\ and
D.\ Y.\ acknowledge financial support from the Access to Major
Research Facilities Program, under the International Science
Linkages Program of the Australian Federal Government. 
Australian
access to the Magellan Telescopes was supported through the Major
National Research Facilities program.  
Observations with the Keck Telescope were made under Gemini exchange
time programs GN-2007B-C-20 and GN-2008A-C-6. 
N.\ C.\ acknowledges financial support for this work through the
Global Networks program of Universit\"at Heidelberg and
Sonderforschungsbereich SFB 881 "The Milky Way System" (subproject
A4) of the German Research Foundation (DFG). 
T.\ C.\ B.\ acknowledges partial funding of this work from grants
PHY 02-16783 and PHY 08-22648: Physics Frontier Center/Joint
Institute for Nuclear Astrophysics (JINA), awarded by the
U.S. National Science Foundation. 
P.\ S.\ B.\ acknowledges support from the Royal Swedish Academy 
of Sciences and the Swedish Research Council. P.\ S.\ B.\ is a 
Royal Swedish Academy of Sciences Research Fellow supported by 
a grant from the Knut and Alice Wallenberg Foundation. 
The authors wish to recognize
and acknowledge the very significant cultural role and reverence that
the summit of Mauna Kea has always had within the indigenous Hawaiian
community. We are most fortunate to have the opportunity to conduct
observations from this mountain. 
Finally, we are pleased to acknowledge support from the
European Southern Observatory's Director's Discretionary Time Program.

\noindent{\it Facilities:} {ATT(DBS); Keck:I(HIRES);
Magellan:Clay(MIKE); VLT:Kueyen(UVES)}


\begin{thebibliography}{64}
\expandafter\ifx\csname natexlab\endcsname\relax\def\natexlab#1{#1}\fi

\bibitem[{{Aoki}(2010)}]{aoki10}
{Aoki}, W. 2010, in IAU Symposium, Vol. 265, IAU Symposium, ed. {K.~Cunha,
  M.~Spite, \& B.~Barbuy}, 111 

\bibitem[{{Aoki} {et~al.}(2007){Aoki}, {Beers}, {Christlieb}, {Norris}, {Ryan},
  \& {Tsangarides}}]{aoki07}
{Aoki}, W., {Beers}, T.~C., {Christlieb}, N., {Norris}, J.~E., {Ryan}, S.~G.,
  \& {Tsangarides}, S. 2007, \apj, 655, 492

\bibitem[{{Aoki} {et~al.}(2006){Aoki}, {Frebel}, {Christlieb}, {Norris},
  {Beers}, {Minezaki}, {Barklem}, {Honda}, {Takada-Hidai}, {Asplund}, {Ryan},
  {Tsangarides}, {Eriksson}, {Steinhauer}, {Deliyannis}, {Nomoto}, {Fujimoto},
  {Ando}, {Yoshii}, \& {Kajino}}]{aoki06}
{Aoki}, W., {Frebel}, A., {Christlieb}, N., {Norris}, J.~E., {Beers}, T.~C.,
  {Minezaki}, T., {Barklem}, P.~S., {Honda}, S., {Takada-Hidai}, M., {Asplund},
  M., {Ryan}, S.~G., {Tsangarides}, S., {Eriksson}, K., {Steinhauer}, A.,
  {Deliyannis}, C.~P., {Nomoto}, K., {Fujimoto}, M.~Y., {Ando}, H., {Yoshii},
  Y., \& {Kajino}, T. 2006, \apj, 639, 897

\bibitem[{{Aoki} {et~al.}(2002){Aoki}, {Norris}, {Ryan}, {Beers}, \&
  {Ando}}]{aoki02}
{Aoki}, W., {Norris}, J.~E., {Ryan}, S.~G., {Beers}, T.~C., \& {Ando}, H. 2002,
  \pasj, 54, 933

\bibitem[{{Asplund}(2005)}]{asplund05}
{Asplund}, M. 2005, \araa, 43, 481

\bibitem[{{Asplund} {et~al.}(2009){Asplund}, {Grevesse}, {Sauval}, \&
  {Scott}}]{asplund09}
{Asplund}, M., {Grevesse}, N., {Sauval}, A.~J., \& {Scott}, P. 2009, \araa, 47,
  481

\bibitem[{{Barklem} {et~al.}(2005){Barklem}, {Christlieb}, {Beers}, {Hill},
  {Bessell}, {Holmberg}, {Marsteller}, {Rossi}, {Zickgraf}, \&
  {Reimers}}]{barklem05}
{Barklem}, P.~S., {Christlieb}, N., {Beers}, T.~C., {Hill}, V., {Bessell},
  M.~S., {Holmberg}, J., {Marsteller}, B., {Rossi}, S., {Zickgraf}, F., \&
  {Reimers}, D. 2005, \aap, 439, 129

\bibitem[{{Beers} \& {Christlieb}(2005)}]{beers05}
{Beers}, T.~C. \& {Christlieb}, N. 2005, \araa, 43, 531

\bibitem[{{Beers} {et~al.}(1985){Beers}, {Preston}, \& {Shectman}}]{beers85}
{Beers}, T.~C., {Preston}, G.~W., \& {Shectman}, S.~A. 1985, \aj, 90, 2089

\bibitem[{{Beers} {et~al.}(1992){Beers}, {Preston}, \& {Shectman}}]{beers92}
---. 1992, \aj, 103, 1987

\bibitem[{{Beers} {et~al.}(1999){Beers}, {Rossi}, {Norris}, {Ryan}, \&
  {Shefler}}]{beers99}
{Beers}, T.~C., {Rossi}, S., {Norris}, J.~E., {Ryan}, S.~G., \& {Shefler}, T.
  1999, \aj, 117, 981

\bibitem[{{Bonifacio} {et~al.}(2007){Bonifacio}, {Molaro}, {Sivarani},
  {Cayrel}, {Spite}, {Spite}, {Plez}, {Andersen}, {Barbuy}, {Beers}, {Depagne},
  {Hill}, {Fran{\c c}ois}, {Nordstr{\"o}m}, \& {Primas}}]{bonifacio07}
{Bonifacio}, P., {Molaro}, P., {Sivarani}, T., {Cayrel}, R., {Spite}, M.,
  {Spite}, F., {Plez}, B., {Andersen}, J., {Barbuy}, B., {Beers}, T.~C.,
  {Depagne}, E., {Hill}, V., {Fran{\c c}ois}, P., {Nordstr{\"o}m}, B., \&
  {Primas}, F. 2007, \aap, 462, 851

\bibitem[{{Bonifacio} {et~al.}(2009){Bonifacio}, {Spite}, {Cayrel}, {Hill},
  {Spite}, {Fran{\c c}ois}, {Plez}, {Ludwig}, {Caffau}, {Molaro}, {Depagne},
  {Andersen}, {Barbuy}, {Beers}, {Nordstr{\"o}m}, \& {Primas}}]{bonifacio09}
{Bonifacio}, P., {Spite}, M., {Cayrel}, R., {Hill}, V., {Spite}, F., {Fran{\c
  c}ois}, P., {Plez}, B., {Ludwig}, H., {Caffau}, E., {Molaro}, P., {Depagne},
  E., {Andersen}, J., {Barbuy}, B., {Beers}, T.~C., {Nordstr{\"o}m}, B., \&
  {Primas}, F. 2009, \aap, 501, 519

\bibitem[{{Bromm} \& {Larson}(2004)}]{bromm04}
{Bromm}, V. \& {Larson}, R.~B. 2004, \araa, 42, 79

\bibitem[{{Caffau} {et~al.}(2011){Caffau}, {Bonifacio}, {Fran{\c c}ois},
  {Sbordone}, {Monaco}, {Spite}, {Spite}, {Ludwig}, {Cayrel}, {Zaggia},
  {Hammer}, {Randich}, {Molaro}, \& {Hill}}]{caffau11}
{Caffau}, E., {Bonifacio}, P., {Fran{\c c}ois}, P., {Sbordone}, L., {Monaco},
  L., {Spite}, M., {Spite}, F., {Ludwig}, H.-G., {Cayrel}, R., {Zaggia}, S.,
  {Hammer}, F., {Randich}, S., {Molaro}, P., \& {Hill}, V. 2011, \nat, 477, 67

\bibitem[{{Carollo} {et~al.}(2012){Carollo}, {Beers}, {Bovy}, {Sivarani},
  {Norris}, {Freeman}, {Aoki}, {Lee}, \& {Kennedy}}]{carollo12}
{Carollo}, D., {Beers}, T.~C., {Bovy}, J., {Sivarani}, T., {Norris}, J.~E.,
  {Freeman}, K.~C., {Aoki}, W., {Lee}, Y.~S., \& {Kennedy}, C.~R. 2012, \apj,
  744, 195

\bibitem[{{Carretta} {et~al.}(2002){Carretta}, {Gratton}, {Cohen}, {Beers}, \&
  {Christlieb}}]{carretta02}
{Carretta}, E., {Gratton}, R., {Cohen}, J.~G., {Beers}, T.~C., \& {Christlieb},
  N. 2002, \aj, 124, 481

\bibitem[{{Castelli} \& {Kurucz}(2003)}]{castelli03}
{Castelli}, F. \& {Kurucz}, R.~L. 2003, in IAU Symp. 210, Modelling of Stellar
  Atmospheres, ed.\ N.\ Piskunov, W.\ W.\ Weiss, \& D.\ F.\ Gray (San
  Francisco, CA: ASP), A20

\bibitem[{{Cayrel} {et~al.}(2004){Cayrel}, {Depagne}, {Spite}, {Hill}, {Spite},
  {Fran{\c c}ois}, {Plez}, {Beers}, {Primas}, {Andersen}, {Barbuy},
  {Bonifacio}, {Molaro}, \& {Nordstr{\"o}m}}]{cayrel04}
{Cayrel}, R., {Depagne}, E., {Spite}, M., {Hill}, V., {Spite}, F., {Fran{\c
  c}ois}, P., {Plez}, B., {Beers}, T., {Primas}, F., {Andersen}, J., {Barbuy},
  B., {Bonifacio}, P., {Molaro}, P., \& {Nordstr{\"o}m}, B. 2004, \aap, 416,
  1117

\bibitem[{{Cescutti} \& {Chiappini}(2010)}]{cescutti10}
{Cescutti}, G. \& {Chiappini}, C. 2010, \aap, 515, 102

\bibitem[{{Christlieb} {et~al.}(2004){Christlieb}, {Gustafsson}, {Korn},
  {Barklem}, {Beers}, {Bessell}, {Karlsson}, \&
  {Mizuno-Wiedner}}]{christlieb04}
{Christlieb}, N., {Gustafsson}, B., {Korn}, A.~J., {Barklem}, P.~S., {Beers},
  T.~C., {Bessell}, M.~S., {Karlsson}, T., \& {Mizuno-Wiedner}, M. 2004, \apj,
  603, 708

\bibitem[{{Cohen} {et~al.}(2002){Cohen}, {Christlieb}, {Beers}, {Gratton}, \&
  {Carretta}}]{cohen02}
{Cohen}, J.~G., {Christlieb}, N., {Beers}, T.~C., {Gratton}, R., \& {Carretta},
  E. 2002, \aj, 124, 470

\bibitem[{{Cohen} {et~al.}(2008){Cohen}, {Christlieb}, {McWilliam}, {Shectman},
  {Thompson}, {Melendez}, {Wisotzki}, \& {Reimers}}]{cohen08}
{Cohen}, J.~G., {Christlieb}, N., {McWilliam}, A., {Shectman}, S., {Thompson},
  I., {Melendez}, J., {Wisotzki}, L., \& {Reimers}, D. 2008, \apj, 672, 320

\bibitem[{{Cohen} {et~al.}(2006){Cohen}, {McWilliam}, {Shectman}, {Thompson},
  {Christlieb}, {Melendez}, {Ramirez}, {Swensson}, \& {Zickgraf}}]{cohen06}
{Cohen}, J.~G., {McWilliam}, A., {Shectman}, S., {Thompson}, I., {Christlieb},
  N., {Melendez}, J., {Ramirez}, S., {Swensson}, A., \& {Zickgraf}, F. 2006,
  \aj, 132, 137

\bibitem[{{Cohen} {et~al.}(2005){Cohen}, {Shectman}, {Thompson}, {McWilliam},
  {Christlieb}, {Melendez}, {Zickgraf}, {Ram{\'{\i}}rez}, \&
  {Swenson}}]{cohen05}
{Cohen}, J.~G., {Shectman}, S., {Thompson}, I., {McWilliam}, A., {Christlieb},
  N., {Melendez}, J., {Zickgraf}, F., {Ram{\'{\i}}rez}, S., \& {Swenson}, A.
  2005, \apjl, 633, L109

\bibitem[{{Frebel} {et~al.}(2006){Frebel}, {Christlieb}, {Norris}, {Beers},
  {Bessell}, {Rhee}, {Fechner}, {Marsteller}, {Rossi}, {Thom}, {Wisotzki}, \&
  {Reimers}}]{frebel06}
{Frebel}, A., {Christlieb}, N., {Norris}, J.~E., {Beers}, T.~C., {Bessell},
  M.~S., {Rhee}, J., {Fechner}, C., {Marsteller}, B., {Rossi}, S., {Thom}, C.,
  {Wisotzki}, L., \& {Reimers}, D. 2006, \apj, 652, 1585

\bibitem[{{Frebel} {et~al.}(2007{\natexlab{a}}){Frebel}, {Johnson}, \&
  {Bromm}}]{frebel07b}
{Frebel}, A., {Johnson}, J.~L., \& {Bromm}, V. 2007{\natexlab{a}}, \mnras, 380,
  L40

\bibitem[{{Frebel} \& {Norris}(2011)}]{frebel11}
{Frebel}, A. \& {Norris}, J.~E. 2011, arXiv:1102.1748 

\bibitem[{{Frebel} {et~al.}(2007{\natexlab{b}}){Frebel}, {Norris}, {Aoki},
  {Honda}, {Bessell}, {Takada-Hidai}, {Beers}, \& {Christlieb}}]{frebel07}
{Frebel}, A., {Norris}, J.~E., {Aoki}, W., {Honda}, S., {Bessell}, M.~S.,
  {Takada-Hidai}, M., {Beers}, T.~C., \& {Christlieb}, N. 2007{\natexlab{b}},
  \apj, 658, 534

\bibitem[Gunn et al.(2006)]{gunn06} Gunn, J.~E., Siegmund, 
W.~A., Mannery, E.~J., et al.\ 2006, \aj, 131, 2332 

\bibitem[{{Hartwick}(1976)}]{hartwick76}
{Hartwick}, F.~D.~A. 1976, \apj, 209, 418

\bibitem[{{Honda} {et~al.}(2004){Honda}, {Aoki}, {Ando}, {Izumiura}, {Kajino},
  {Kambe}, {Kawanomoto}, {Noguchi}, {Okita}, {Sadakane}, {Sato},
  {Takada-Hidai}, {Takeda}, {Watanabe}, {Beers}, {Norris}, \& {Ryan}}]{honda04}
{Honda}, S., {Aoki}, W., {Ando}, H., {Izumiura}, H., {Kajino}, T., {Kambe}, E.,
  {Kawanomoto}, S., {Noguchi}, K., {Okita}, K., {Sadakane}, K., {Sato}, B.,
  {Takada-Hidai}, M., {Takeda}, Y., {Watanabe}, E., {Beers}, T.~C., {Norris},
  J.~E., \& {Ryan}, S.~G. 2004, \apjs, 152, 113

\bibitem[{{Izzard} {et~al.}(2009){Izzard}, {Glebbeek}, {Stancliffe}, \&
  {Pols}}]{izzard09}
{Izzard}, R.~G., {Glebbeek}, E., {Stancliffe}, R.~J., \& {Pols}, O.~R. 2009,
  \aap, 508, 1359

\bibitem[{{Karlsson}(2006)}]{karlsson06}
{Karlsson}, T. 2006, \apjl, 641, L41

\bibitem[{{Kobayashi} {et~al.}(2006){Kobayashi}, {Umeda}, {Nomoto}, {Tominaga},
  \& {Ohkubo}}]{kobayashi06}
{Kobayashi}, C., {Umeda}, H., {Nomoto}, K., {Tominaga}, N., \& {Ohkubo}, T.
  2006, \apj, 653, 1145

\bibitem[{{Komiya} {et~al.}(2007){Komiya}, {Suda}, {Minaguchi}, {Shigeyama},
  {Aoki}, \& {Fujimoto}}]{komiya07}
{Komiya}, Y., {Suda}, T., {Minaguchi}, H., {Shigeyama}, T., {Aoki}, W., \&
  {Fujimoto}, M.~Y. 2007, \apj, 658, 367

\bibitem[{{Lai} {et~al.}(2008){Lai}, {Bolte}, {Johnson}, {Lucatello}, {Heger},
  \& {Woosley}}]{lai08}
{Lai}, D.~K., {Bolte}, M., {Johnson}, J.~A., {Lucatello}, S., {Heger}, A., \&
  {Woosley}, S.~E. 2008, \apj, 681, 1524

\bibitem[{{Laird} {et~al.}(1988){Laird}, {Carney}, {Rupen}, \&
  {Latham}}]{laird88}
{Laird}, J.~B., {Carney}, B.~W., {Rupen}, M.~P., \& {Latham}, D.~W. 1988, \aj,
  96, 1908

\bibitem[{{Li} {et~al.}(2010){Li}, {Christlieb}, {Sch{\"o}rck}, {Norris},
  {Bessell}, {Yong}, {Beers}, {Lee}, {Frebel}, \& {Zhao}}]{li10}
{Li}, H.~N., {Christlieb}, N., {Sch{\"o}rck}, T., {Norris}, J.~E., {Bessell},
  M.~S., {Yong}, D., {Beers}, T.~C., {Lee}, Y.~S., {Frebel}, A., \& {Zhao}, G.
  2010, \aap, 521, A10

\bibitem[{{Lucatello} {et~al.}(2006){Lucatello}, {Beers}, {Christlieb},
  {Barklem}, {Rossi}, {Marsteller}, {Sivarani}, \& {Lee}}]{lucatello06}
{Lucatello}, S., {Beers}, T.~C., {Christlieb}, N., {Barklem}, P.~S., {Rossi},
  S., {Marsteller}, B., {Sivarani}, T., \& {Lee}, Y.~S. 2006, \apjl, 652, L37

\bibitem[{{Lucatello} {et~al.}(2005){Lucatello}, {Gratton}, {Beers}, \&
  {Carretta}}]{lucatello05}
{Lucatello}, S., {Gratton}, R.~G., {Beers}, T.~C., \& {Carretta}, E. 2005,
  \apj, 625, 833

\bibitem[{{Norris}(1999)}]{norris99}
{Norris}, J.~E. 1999, in Astronomical Society of the Pacific Conference Series,
  Vol. 165, The Third Stromlo Symposium: The Galactic Halo, ed. {B.~K.~Gibson,
  R.~S.~Axelrod, \& M.~E.~Putman}, 213

\bibitem[{{Norris} {et~al.}(2007){Norris}, {Christlieb}, {Korn}, {Eriksson},
  {Bessell}, {Beers}, {Wisotzki}, \& {Reimers}}]{norris07}
{Norris}, J.~E., {Christlieb}, N., {Korn}, A.~J., {Eriksson}, K., {Bessell},
  M.~S., {Beers}, T.~C., {Wisotzki}, L., \& {Reimers}, D. 2007, \apj, 670, 774

\bibitem[{{Norris} {et~al.}(2001){Norris}, {Ryan}, \& {Beers}}]{norris01}
{Norris}, J.~E., {Ryan}, S.~G., \& {Beers}, T.~C. 2001, \apj, 561, 1034

\bibitem[{{Pagel} \& {Patchett}(1975)}]{pagel75}
{Pagel}, B.~E.~J. \& {Patchett}, B.~E. 1975, \mnras, 172, 13

\bibitem[{{Prantzos}(2008)}]{prantzos08}
{Prantzos}, N. 2008, \aap, 489, 525

\bibitem[{{Ryan} \& {Norris}(1991)}]{ryan91}
{Ryan}, S.~G. \& {Norris}, J.~E. 1991, \aj, 101, 1865

\bibitem[{{Salvadori} {et~al.}(2007){Salvadori}, {Schneider}, \&
  {Ferrara}}]{salvadori07}
{Salvadori}, S., {Schneider}, R., \& {Ferrara}, A. 2007, \mnras, 381, 647

\bibitem[{{Schmidt}(1963)}]{schmidt63}
{Schmidt}, M. 1963, \apj, 137, 758

\bibitem[{{Sch{\"o}rck} {et~al.}(2009){Sch{\"o}rck}, {Christlieb}, {Cohen},
  {Beers}, {Shectman}, {Thompson}, {McWilliam}, {Bessell}, {Norris},
  {Mel{\'e}ndez}, {Ram{\'{\i}}rez}, {Haynes}, {Cass}, {Hartley}, {Russell},
  {Watson}, {Zickgraf}, {Behnke}, {Fechner}, {Fuhrmeister}, {Barklem},
  {Edvardsson}, {Frebel}, {Wisotzki}, \& {Reimers}}]{schorck09}
{Sch{\"o}rck}, T., {Christlieb}, N., {Cohen}, J.~G., {Beers}, T.~C.,
  {Shectman}, S., {Thompson}, I., {McWilliam}, A., {Bessell}, M.~S., {Norris},
  J.~E., {Mel{\'e}ndez}, J., {Ram{\'{\i}}rez}, S., {Haynes}, D., {Cass}, P.,
  {Hartley}, M., {Russell}, K., {Watson}, F., {Zickgraf}, F., {Behnke}, B.,
  {Fechner}, C., {Fuhrmeister}, B., {Barklem}, P.~S., {Edvardsson}, B.,
  {Frebel}, A., {Wisotzki}, L., \& {Reimers}, D. 2009, \aap, 507, 817

\bibitem[{{Searle} \& {Sargent}(1972)}]{searle72}
{Searle}, L. \& {Sargent}, W.~L.~W. 1972, \apj, 173, 25

\bibitem[{Shimazaki \& Shinomoto(2007)}]{shimazaki07}
Shimazaki, H. \& Shinomoto, S. 2007, Neural Computation, 19, 1503

\bibitem[{{Siegel}(1956)}]{siegel56}
{Siegel}, S. 1956, {Nonparametric statistics for the behavioral sciences (Tokyo: McGraw-Hill Kogakusku)}

\bibitem[{{Skrutskie} {et~al.}(2006){Skrutskie}, {Cutri}, {Stiening},
  {Weinberg}, {Schneider}, {Carpenter}, {Beichman}, {Capps}, {Chester},
  {Elias}, {Huchra}, {Liebert}, {Lonsdale}, {Monet}, {Price}, {Seitzer},
  {Jarrett}, {Kirkpatrick}, {Gizis}, {Howard}, {Evans}, {Fowler}, {Fullmer},
  {Hurt}, {Light}, {Kopan}, {Marsh}, {McCallon}, {Tam}, {Van Dyk}, \&
  {Wheelock}}]{2mass}
{Skrutskie}, M.~F., {Cutri}, R.~M., {Stiening}, R., {Weinberg}, M.~D.,
  {Schneider}, S., {Carpenter}, J.~M., {Beichman}, C., {Capps}, R., {Chester},
  T., {Elias}, J., {Huchra}, J., {Liebert}, J., {Lonsdale}, C., {Monet}, D.~G.,
  {Price}, S., {Seitzer}, P., {Jarrett}, T., {Kirkpatrick}, J.~D., {Gizis},
  J.~E., {Howard}, E., {Evans}, T., {Fowler}, J., {Fullmer}, L., {Hurt}, R.,
  {Light}, R., {Kopan}, E.~L., {Marsh}, K.~A., {McCallon}, H.~L., {Tam}, R.,
  {Van Dyk}, S., \& {Wheelock}, S. 2006, \aj, 131, 1163

\bibitem[{{Sneden}(1973)}]{moog}
{Sneden}, C. 1973, \apj, 184, 839

\bibitem[{{Sneden} {et~al.}(2008){Sneden}, {Cowan}, \& {Gallino}}]{sneden08}
{Sneden}, C., {Cowan}, J.~J., \& {Gallino}, R. 2008, \araa, 46, 241

\bibitem[{{Sobeck} {et~al.}(2011){Sobeck}, {Kraft}, {Sneden}, {Preston},
  {Cowan}, {Smith}, {Thompson}, {Shectman}, \& {Burley}}]{sobeck11}
{Sobeck}, J.~S., {Kraft}, R.~P., {Sneden}, C., {Preston}, G.~W., {Cowan},
  J.~J., {Smith}, G.~H., {Thompson}, I.~B., {Shectman}, S.~A., \& {Burley},
  G.~S. 2011, \aj, 141, 175

\bibitem[{{Spite} {et~al.}(2005){Spite}, {Cayrel}, {Plez}, {Hill}, {Spite},
  {Depagne}, {Fran{\c c}ois}, {Bonifacio}, {Barbuy}, {Beers}, {Andersen},
  {Molaro}, {Nordstr{\"o}m}, \& {Primas}}]{spite05}
{Spite}, M., {Cayrel}, R., {Plez}, B., {Hill}, V., {Spite}, F., {Depagne}, E.,
  {Fran{\c c}ois}, P., {Bonifacio}, P., {Barbuy}, B., {Beers}, T., {Andersen},
  J., {Molaro}, P., {Nordstr{\"o}m}, B., \& {Primas}, F. 2005, \aap, 430, 655

\bibitem[{{Suda} {et~al.}(2008){Suda}, {Katsuta}, {Yamada}, {Suwa}, {Ishizuka},
  {Komiya}, {Sorai}, {Aikawa}, \& {Fujimoto}}]{saga}
{Suda}, T., {Katsuta}, Y., {Yamada}, S., {Suwa}, T., {Ishizuka}, C., {Komiya},
  Y., {Sorai}, K., {Aikawa}, M., \& {Fujimoto}, M.~Y. 2008, \pasj, 60, 1159

\bibitem[{{Wisotzki} {et~al.}(1996){Wisotzki}, {Koehler}, {Groote}, \&
  {Reimers}}]{hes}
{Wisotzki}, L., {Koehler}, T., {Groote}, D., \& {Reimers}, D. 1996, \aaps, 115,
  227

\bibitem[York et al.(2000)]{york00} York, D.~G., Adelman, J., 
Anderson, J.~E., Jr., et al.\ 2000, \aj, 120, 1579 

\end{thebibliography}
\end{document}